\documentstyle[11pt]{article}
\def\thefootnote{\dag}

\newcommand{\overl}{\overline}
\newcommand{\be}{\begin{equation}}
\newcommand{\ee}{\end{equation}}
\newcommand{\bea}{\begin{eqnarray}}
\newcommand{\eea}{\end{eqnarray}}
\newcommand{\dy}{\displaystyle}
\newcommand{\PL}{{\it Physics Letters}}
\newcommand{\NC}{{\it Il Nuovo Cimento}}
\newcommand{\PRL}{{\it Physical Review Letters}}
\newcommand{\PR}{{\it Physical Review}}
\newcommand{\ZfP}{{\it Zeitschrift f\"ur Physik}}
\newcommand{\PTP}{{\it Progress of Theoretical Physics}}
\begin{document}
\thispagestyle{empty}
\baselineskip=20pt
\hfill{
\begin{tabular}{l}
DSF$-$93/12\\
INFN$-$NA$-$IV$-$93/12\\
To~appear~in~\ZfP~{\bf C}
\end{tabular}}

\bigskip\bigskip

\baselineskip=30pt

\begin{center}
{\huge
{\bf  Rare $B$-decays and Heavy to Light Semileptonic Transitions
in the Isgur and Wise Limit}}
\end{center}

\vspace{2cm}

\baselineskip=20pt

\begin{center}
{\large
{\bf Pietro Santorelli\\}}

\bigskip\bigskip

{\it{\footnotesize
\noindent
Dipartimento di Scienze Fisiche, Universit\`a di Napoli,
Mostra d'Oltremare Pad. 19, I-80125 Napoli Italy\\
\noindent
INFN, Sezione di Napoli,
Mostra d'Oltremare Pad. 20, I-80125 Napoli Italy}}
\end{center}

\vspace{1cm}


\vspace{1cm}

\begin{center}
{\bf Abstract}\\
\noindent
{}From the experimental branching ratios for
$B^-\rightarrow\rho^0~l^-\bar\nu_l$
and $D^+ \rightarrow {\overl K}^{*0}({\overl K}^0)~e^+ \nu_e$ one finds, in the
Heavy Quark Limit of $HQET$, $~~|V_{bu}|=(8.1\pm 1.7)\times 10^{-3}$, larger
but consistent with the actual quoted range $(2~-~7)\times 10^{-3}$. In the
same framework one predicts for $R(B\rightarrow K^*\gamma)=(~2~ \pm~
2~)~10^{-2}$.
\end{center}

\vspace{2cm}

\centerline{{\bf PACS}: 13.20.Jf, 11.30.Ly}

\noindent

\newpage

The study of the Cabibbo-Kobayashi-Maskawa \cite{CKM} suppressed decays
$B^-\rightarrow\rho^0~e^-\bar\nu_e$, interesting in itself for the
determination of $|V_{bu}|$ \cite{1}, has been recently
related by the spin-flavour symmetries of the $HQET$
\cite{2} (in the Heavy Quark Limit) to the rare
$B$-decays \cite{3}\cite{4}. In such a way the predictions for the branching
ratio of the decay $B\rightarrow K^*\gamma$, which provide a test of the
Standard Model \cite{trampetic}, depend strongly on the value of $|V_{bu}|$,
for which the experimental data about the $b\rightarrow u~e~\bar\nu_e$
(inclusive ed exclusive) decays give an information depending also from
the theoretical approach followed to evaluate the corresponding amplitudes.

Here we shall reach rather firm conclusions by following the suggestion of
N. Isgur and M.B. Wise of relating the involved form factors
by flavour symmetry \cite{6}.

The $\bar B\rightarrow\rho$ semileptonic decays are chosen rather than
the decays with $\pi$ in the final state, because in this last case the
upper limit of the invariant mass of the final leptons
is very near to the $B^*$ resonance (which is expected equal
to $m_B$ in the $HQL$); as a consequence the pole
of the $B^*$ dominates the spectrum in that region so that the
prediction of $HQET$ fails near to the no-recoil point \cite{IWB*}.

In the first section we review the spectrum $B\rightarrow V~l~\nu_l$ and give
the involved form factors. In the second we get
$|V_{bu}|$ by comparing experiment with the theoretical predictions. In the
third are discussed the rare $B$ decays in the $HQL$.

\vspace{1.5cm}

\noindent
{\Large {\bf 1.}$~~$} The spectrum in the invariant mass $q^2$ of the
lepton pair for the semileptonic decays of a heavy meson
with one vector meson in the final state ($H_j\rightarrow V_k~ l~\nu_l$)
is given by e.g. in \cite{7} (neglecting lepton masses):
\bea
\frac{d\Gamma(H_j\rightarrow V_k~e~\nu_e)}{dq^2} & = &
\frac{G_F^2|V_{jk}|^2}{192\pi^3m_H^3}\sqrt{\lambda(m_H^2,m_V^2,q^2)}
\nonumber\\
&\cdot &\left\{|A^{(jk)}_1(q^2)|^2\left[2(m_H+m_V)^2q^2+
\frac{(m_H+m_V)^2}{4m_V^2}(m_H^2-m_V^2-q^2)^2\right]\right.\nonumber\\
& + &\left. |A^{(jk)}_2(q^2)|^2~
\frac{\lambda^2(m_H^2,m_V^2,q^2)}{4m_V^2(m_H+m_V)^2}
\right.\nonumber\\
& - & \left.A^{(jk)}_1(q^2)A^{(jk)}_2(q^2)~
\lambda(m_H^2,m_V^2,q^2)\frac{(m_H^2-m_V^2-q^2)}{2m_V^2}\right.
\nonumber\\
& + &\left.
|V^{(jk)}(q^2)|^2\frac{2q^2}{(m_H+m_V)^2}~\lambda(m_H^2,m_V^2,q^2)\right\}~;
\eea
where
\bea
\lambda(x,y,z)& = & x^2+y^2+z^2-2(xy+xz+yz)~,\\
q^2 & = & (p_H-p_{V})^2~.
\eea

The form factors of the weak currents for an initial $\overl B$ meson are:
\bea
\langle V_j(\varepsilon ,p_j)|(A^{\mu})^b_j|{\overl B}(p)\rangle & = &
(m_B+m_V)A_1^{(bj)}(q^2)\left(\varepsilon^{*\mu}-
\frac{\varepsilon^*\cdot q}{q^2}q^{\mu}\right)
\nonumber\\
& - & A_2^{(bj)}(q^2)\frac{\varepsilon^*\cdot q}{m_V+m_B}
\left(p_j^{\mu}+p^{\mu}-\frac{m_B^2-m_V^2}{q^2}q^{\mu}\right)
\nonumber\\
& + & 2m_VA_0^{(bj)}(q^2)\frac{\varepsilon^*\cdot q}{q^2}q^{\mu}~,
\\
\langle V_j(\varepsilon ,p_j)|(V^{\mu})^b_j|{\overl B}(p)\rangle & = &
2iV^{(bj)}(q^2)\frac{ \varepsilon^{\mu}_{~~\nu\rho\sigma}p^{\nu}p^{\rho}_j
\varepsilon
^{*\sigma} }{m_B+m_V}~.
\eea
and are related, by $HQET$ (leading order), to the corresponding ones
for the process $D\rightarrow V~ l~\nu_l$ if $V$ is a light vector meson.

N. Isgur  and M.B. Wise \cite{6} found as a consequence of flavour symmetry
the relations between the weak form factors for $\bar B\rightarrow K^*$ and
$D\rightarrow K^*$, which, for the our parameterization, imply:
\bea
A_1^{(bj)}(q^2_B)& = &C_{bc}\left(\frac{m_D+m_V}{m_B+m_V}\right)
\sqrt{\frac{m_B}{m_D}}A_1^{(cj)}(q^2_D)~,\\
\nonumber\\
A_2^{(bj)}(q^2_B)& = &\frac{C_{bc}}{2}
\sqrt{\left(\frac{m_D}{m_B}\right)^3}
\left(\frac{m_B+m_V}{q^2_D}\right)\left\{
y_2(m_D+m_V)A_1^{(cj)}(q^2_D)
\right.\nonumber\\
& + &\left.A_2^{(cj)}(q^2_D)\left[
\frac{y_1q^2_D}{m_D+m_V}-y_2(m_D-m_V)
\right]-2m_Vy_2A_0^{(cj)}(q^2_D)\right\}~,\\
\nonumber\\
A_0^{(bj)}(q^2_B)& = &\frac{C_{bc}}{4m_V}
\sqrt{\left(\frac{m_D}{m_B}\right)^3}\left\{
\frac{2m_Vx_2}{q^2_D}A_0^{(cj)}(q^2_D)+
\left[2\left(\frac{m_B}{m_D}\right)^2-\frac{x_2}{q^2_D}\right](m_D+m_V)
A_1^{(cj)}(q^2_D)
\right.\nonumber\\
& + &\left.
\left[(m_D-m_V)\frac{x_2}{q^2_D}-\frac{x_1}{m_D+m_V}\right]A_2^{(cj)}(q^2_D)
\right\}~,\\
\nonumber\\
V^{(bj)}(q^2_B)& = &C_{bc}\frac{m_B+m_V}{m_D+m_V}
\sqrt{\frac{m_D}{m_B}}V^{(cj)}(q^2_D)~,\\
\nonumber
\eea
where
\be
\begin{array}{c}
\\
\dy q^2_B=(m_Bv-p_V)^2~~~~ q^2_D=(m_Dv-p_V)^2~~~~~~v^2=1~,\\
\\
\dy y_1=\frac{(m_B+m_D)}{m_D} ~~~ y_2=\frac{(m_D-m_B)}{m_D}~,\\
\\
\dy x_1=y_2q^2_B+y_1(m_B^2-m_V^2) ~~ x_2=y_1q^2_B+y_2(m_B^2-m_V^2)~,\\
\\
\dy C_{bc}\cong\left(\frac{\alpha_s(m_b)}{\alpha_s(m_c)}\right)^{-6/25}
\cong 1.1~.
\end{array}
\ee

It is worth noticing that the $q^2_D$ values corresponding to the physical
region for $q^2_B$ are not located in the $q^2$ allowed range for
$D\rightarrow K^*$ semileptonic decays.

The symmetry breaking corrections
to equations (6)-(9) are proportional in the full kinematical range to
$\alpha_s m_L/m_H$, where $\alpha_s$ is the coupling constant of $QCD$ and
$m_L$ is the mass of light quark coupled to $b$ by the weak current;
for the case considered here $V\equiv\rho$,$~~m_L=m_u$ and
therefore we may safely neglect these corrections \cite{5}.

\vspace{1.5cm}

\noindent
{\Large {\bf 2.}$~~$}From the experimental data on the semileptonic
$D^+\rightarrow {\overl K}^0({\overl K}^{0*})~e^+~\nu_e$ decays,
within the pole approximation
for the form factors, the $E691$ Collaboration \cite{8} found the following
values for their residua:
\be
\begin{array}{c}
\dy A_0^{(cs)}(0)=0.71 \pm 0.16~,~~~A_2^{(cs)}(0)=0.00 \pm 0.22~,
{}~~~V^{(cs)}(0)=0.90 \pm 0.32~,\\
\\
\dy A_1^{(cs)}(0) \equiv \frac{1}{m_D+m_{K^*}}
\left\{ 2m_{K^*}A_0^{(cs)}(0)+(m_D-m_{K^*})A_2^{(cs)}(0) \right\}~.
\\
\end{array}
\ee

{}From $SU(3)$ invariance we may identify the residua for
$D^0\rightarrow K^{*-}$ with ones for $D^0\rightarrow \rho^-$ weak form factors
and from equations (6)-(9) and the values given in equation (11)
one may predict the form of the spectrum for
\[\frac{d\Gamma(B\rightarrow\rho~e~\nu_e)}{dq^2}~,\]
and the rate
\be
\frac{1}{|V_{bu}|^2}\Gamma(B^- \rightarrow \rho^0~l^-~\bar\nu_l) =
0.80\times 10^{-11}~~GeV~.
\ee

{}From the measured branching ratio
$Br(B^-\rightarrow\rho^0~l^-~\bar\nu_l)=(10.3 \pm 3.6 \pm 2.5)\times 10^{-4}$
\cite{Argus-b-u} one obtains
(neglecting the errors on values of the form factors in equation (11))
\be
|V_{bu}| = \sqrt{\frac{Br(B^-\rightarrow\rho^0~l^-\bar\nu_l)|_{exp.}}
{\Gamma(B^-\rightarrow\rho^0l^-\bar\nu_l)/\tau_{B}}}
=(8.1 \pm 1.7)\times 10^{-3}
\ee
larger but still consistent, within the experimental uncertainties, with the
value found from the experimental information on
$|V_{bu}/V_{bc}|=0.10\pm 0.03$ \cite{1}\cite{vbuvbc} and for the value
$|V_{bc}|=0.043\pm 0.003$ obtained from the study of semileptonic
$B$-decays in the Heavy Quark Limit \cite{BSant}:
\be
|V_{bu}|=(4.3\pm 1.3)\times 10^{-3}~.
\ee

The spectrum predicted is described in figure 1\footnote
{If we ignore the location of $B^*$ resonance and calculate the
spectrum and the rate of $B\rightarrow\pi$ semileptonic decay we obtain a
linear dependance of $d\Gamma/dq^2$ by the $q^2$. The result is very
similar to the one obtained by J.G. K\"{o}rner and G.A. Schuler \cite{KS}.}.

\vspace{1.5cm}

{\Large {\bf 3.}$~~$} A high precision prediction about the rate of
the rare decay $B\rightarrow K^*\gamma$ \cite{bk*g}, which is induced at
one loop in the Standard Model, is very important, since the discrepancy
between theory and experiment would be indirect evidence of new physics.

By relating $B\rightarrow K^*\gamma$ to the semileptonic
$\bar B\rightarrow\rho$ decay, P.J. O'Donnell and H.K.K. Tung \cite{4}
obtained for the ratio
\be
R(B\rightarrow K^*\gamma)=\frac{\Gamma(B\rightarrow K^*\gamma)}
{\Gamma(b\rightarrow s\gamma)}=\frac{m_b^3}{(m_b^2-m_s^2)^3}
\frac{(m_B^2-m_{K^*}^2)^3}{m_B^3}\frac{1}{2}\left\{|F_1(0)|^2+
4|F_2(0)|^2\right\}
\ee
the prediction, which differs from the one deduced by us
\def\thefootnote{\dag}
\footnote{I am indebted to Patrick O'Donnell and
Humphrey Tung for a clarifying communication.}
\be
 R(B\rightarrow K^*\gamma)\left(\left.
\frac {d\Gamma(\bar B\rightarrow \rho l\bar\nu_l)}{dq^2}
\right|_{q^2=0}\right)^{-1}=
\frac{192\pi^3}{G_F^2} \frac{1}{|V_{bu}|^2}
\frac{(m_B^2-m_{K^*}^2)^3}{(m_B^2-m_{\rho}^2)^3}
\frac{m_b^3}{(m_b^2-m_s^2)^3}|{\cal I}|^2~,
\ee
because we invoke $SU(3)_{u,d,s}$ symmetry for $V^{(bj)}(0)$ rather than
for $T_1^{B\rightarrow V_j}(0)=V^{(bj)}(0)/(m_B+m_{V_j})$ as in \cite{4}.

Note that $|{\cal I}|=1$ in the $HQL$ and
the corrections to the prediction of $HQET$ in relating the form factor in
equations (4)-(5) (with $j=u~$or$~s$) to the one of the matrix element
$\langle V_j|{\overl q}_j\sigma_{\mu\nu}q^{\nu}b_R |\bar B\rangle$
are expected to be small (cfr. \cite{4}) and we neglect them.\\
In effect the conclusions of the paper of P.J. O'Donnell and H.K.K.
Tung are that
the corrections to the assumption that $K^*$ and $\rho$ are heavy mesons are
negligible (at least to relate the form factors of the matrix elements
of the weak currents and the $\bar s\sigma_{\mu\nu}b_R$
and $\bar u\sigma_{\mu\nu}b_R$ operators respectively at $q^2=0$).
The universality of the Isgur-Wise function, in principle, allow us to
give, without relating the ratio $R$ to the spectrum of
$B\rightarrow\rho~l~\nu_l$, the $R(B\rightarrow K^*\gamma)$ in terms of the
Isgur-Wise function $\xi(w^2)$ extracted, for example, by the experimental
data on charmed semileptonic $B$-decays (cfr. for example \cite{BSant}).
In such case the following relations hold
\def\thefootnote{\diamond}
\footnote{The first relation is
more general. For example it is necessary to avoid an unphysical pole in
$q^2=0$ for the $h(q^2)$ and $g_-(q^2)$ form factors introduced in \cite{6}.}
\be
\begin{array}{c}
F_1(0)=2F_2(0)~,\\
\\
\dy F_1(q^2)=\frac{(m_B+m_{K^*})}{2\sqrt{m_Bm_{K^*}}}\xi(w^2(q^2))~.
\\
\end{array}
\ee

But the value
\be
w^2(q^2=0) \equiv \left.\left(\frac{p_B}{m_B}-\frac{p_{K^*}}{m_{K^*}}\right)^2
\right|_{q^2=0}
=-\frac{(m_B-m_{K^*})^2}{m_Bm_{K^*}}
\ee
is too far from the physical range of $\overline{B}\rightarrow D^{(*)}$
semileptonic decays and the predictions depend strongly on
the behaviour assumed for $\xi(w^2)$
\def\thefootnote{\ddag}
\footnote{From the data on $D\rightarrow K(K^*)l^-\bar\nu_l$ decays
A. Ali and T. Mannel \cite{AliMannel}
extracted the values $w_0=1.8$ and $\beta=0.25$
respectively for the pole and exponential
parameterization for $\xi(w^2)$.}.

A more reliable prediction for the ratio $R$ from equation (16)
is obtained by relating the spectra of $B\rightarrow\rho~l~\nu_l$ and
$D\rightarrow K^*$ semileptonic decays.

Following the hypothesis of section 2 about the residua of the involved
form factors, $SU(2)$ flavour symmetry and (cfr. (1))
\bea
\left.\frac{d\Gamma(\bar B\rightarrow\rho~l~\bar\nu_l)}{dq^2}\right|_{q^2=0}
& = &
\frac{G_F^2|V_{bu}|^2}{192\pi^3m_B^3}\frac{(m^2_B-m^2_{\rho})^3}{4m_{\rho}^2}
\left|(m_B+m_{\rho})A_1^{(bu)}(0)-(m_B-m_{\rho})A_2^{(bu)}(0)\right|^2
\nonumber\\
& = &
\frac{G_F^2|V_{bu}|^2}{192\pi^3}\frac{(m^2_B-m^2_{\rho})^3}{m_{B}^3}
\cdot\left|A_0^{(bu)}(0)\right|^2~,
\eea
from the equations (16),(8) (or (6) and (7)) and (11) we derive
\be
R(B\rightarrow K^*\gamma)=
\frac{m_b^3}{(m_b^2-m_s^2)^3}
\frac{(m_B^2-m^2_{K^*})^3}{m_B^3}
\cdot\left|A_0^{(bu)}(0)\right|^2
\ee
giving ($m_b=5~GeV$ and $m_s=0.55~GeV$)
\be
R(B\rightarrow K^*\gamma)=(35 \pm 28) \times 10^{-2}~.
\ee

The central value is very near to the one given in
\cite{AliMannel} in the polar approximation for $\xi(w^2)$. The error
quoted in (21) depends on the large error in the determination of
$A^{(cs)}_2(0)$.

Obviously one expects the assumption that $K^*$ and $\rho$ are
heavy less reliable than $HQL$ for $b$ and $c$ quarks;
thus the previous result can be modified
by taking $\cal I$ from ref. \cite{4}:
\be
R(B\rightarrow K^*\gamma)=(35 \pm 28)\times 10^{-2} ~\cdot~ |{\cal I}|^2
=\left\{\begin{array}{c}
(42 \pm 33)\times 10^{-2}~~~for~~~{\cal I}=1.09\\
\\
(49 \pm 39)\times 10^{-2}~~~for~~~{\cal I}=1.18\\
\end{array}\right.
\ee

It is worth recalling that the static limit for $b$
($\gamma_0b=b$) and $SU(2)_{bc}$ heavy flavour symmetry imply
\bea
R(B\rightarrow K^*\gamma) & = & \frac{C^2_{bc}m_b^3}{(m_b^2-m_s^2)^3}
\frac{(m_B^2-m_{K^*}^2)^3}{4m_B^4m_D}
\left|
(m_D+m_{K^*})A_1^{(cs)}(q^2_D)-\left(\frac{m_B^2-m_{K^*}^2}{m_D+m_{K^*}}
\right)\frac{m_D}{m_B}V^{(cs)}(q^2_D)\right|^2
\nonumber\\
\nonumber\\
& = & (~2 \pm 2~)~\cdot~10^{-2}~.
\eea
The result is substantially equivalent to the one dictated by $HQL$ for
$b$, $c$ and $s$ and a wave function model for the
Isgur-Wise function \cite{AOM}.

\vspace{1cm}

We derived, in the $HQL$ for $b$ and $c$, the spectrum predicted for
$B\rightarrow \rho l\nu_l$ and, comparing theory and experiment,
we give $|V_{bu}|$.\\
Also relating $R(B\rightarrow K^*\gamma)$ to the
$\dy \frac{d\Gamma(B\rightarrow \rho l\nu_l)}{dq^2}$ at $q^2=0$ we obtained
the ratio $R$ in terms of $A^{(bu)}_1(0)$ and $A^{(bu)}_2(0)$
(or $A^{(bu)}_0(0)$), estimated by
extrapolation from the corresponding
form factors for $D\rightarrow K^*$ semileptonic decay.\\
We are more confident on the prediction for $R$
coming from the static limit for $b$ and $SU(2)_{bc}$ heavy flavour
symmetry.

A more precise determination of $c\rightarrow s$ weak form
factors is needed for a more precise evaluation for
$R(B\rightarrow K^*\gamma)$.

\vspace{1cm}

\centerline{\bf Acknowledgement}

I am indebted to Professor Franco Buccella for many enlightening
discussions.

\newpage

{\Large\centerline{Figure Caption}}

{\bf Figure 1.}
We report the spectrum predicted for the decay
$B^-\rightarrow \rho^0 l^- \bar\nu_l$.

\end{document}